\RequirePackage{fix-cm}
\documentclass[a4paper,10pt]{article}
\usepackage{datetime}
\usepackage{graphicx}
\usepackage{mathptmx} 
\usepackage[T1]{fontenc}
\usepackage[utf8]{inputenc}
\usepackage{xcolor}
\usepackage[nodisplayskipstretch]{setspace}
\usepackage{mathrsfs,amsmath}
\usepackage{amsbsy}
\usepackage{bbold}
\usepackage{caption}
\usepackage{color}

\title{Numerical simulation of biofilm formation in a microchannel}
\author{David Landa-Marb\'an${}^{1}$, Iuliu Sorin Pop${}^{1,2}$, Kundan  Kumar${}^1$, \and and Florin A. Radu${}^1$}
\date{}
\begin{document}
\maketitle
\noindent ${}^1$ Department of Mathematics, Faculty of Mathematics and Natural Sciences, University of Bergen, All\'egaten 41, P.O. Box 7803, 5020 Bergen, Norway.\\[5pt]
${}^2$ Faculty of Sciences, Hasselt University, Campus Diepenbeek, Agoralaan building D, BE3590 Diepenbeek, Belgium.\\[5pt]
Corresponding author: David Landa-Marb\'an (E-mail: David.Marban@uib.no)
\begin{abstract}
\noindent The focus of this paper is the numerical solution of a mathematical model for the growth of a permeable biofilm in a microchannel. The model includes water flux inside the biofilm, different biofilm components, and shear stress on the biofilm-water interface. To solve the resulting highly coupled system of model equations, we propose a splitting algorithm. The Arbitrary Lagrangian Eulerian (ALE) method is used to track the biofilm-water interface. Numerical simulations are performed using physical parameters from the existing literature. Our computations show the effect of biofilm permeability on the nutrient transport and on its growth. 
\end{abstract}
\section{Model equations}
Oil is one of the principal energy resources. Therefore it is essential to develop efficient extraction techniques with minimal environmental impact. One good candidate is the bio-plug, were bacteria are brought in the high permeable zones. The bacterial growth reduces the permeabilities of these zones, so we can reach and recover the oil in the less permeable zones. Our motivation is to develop mathematical models that have higher fidelity to the experiments and thus describe this mechanism in a better manner. 

A biofilm is an assemblage of surface-associated microbial cells that is enclosed in an extracellular polymeric substance matrix (EPS) \cite{Donlan:Article:2002}. The proportion of EPS in biofilms can comprise between 50-90$\%$ of the total organic matter \cite{Donlan:Article:2002,Vu:Article:2009}. Since water is the largest component of the biofilm \cite{Flemming:Article:2010}, it is important to consider the water flux inside the biofilm in our model. In this section, we describe the model equations for biofilm formation in a strip considering a permeable biofilm \cite{Landa-Marban1:Article:2017}. The mathematical model considered here follows ideas from \cite{Alpkvist:Article:2007,Schulz:Article:2017,vanNoorden:Article:2010}.

\begin{figure}
\includegraphics[scale=.68]{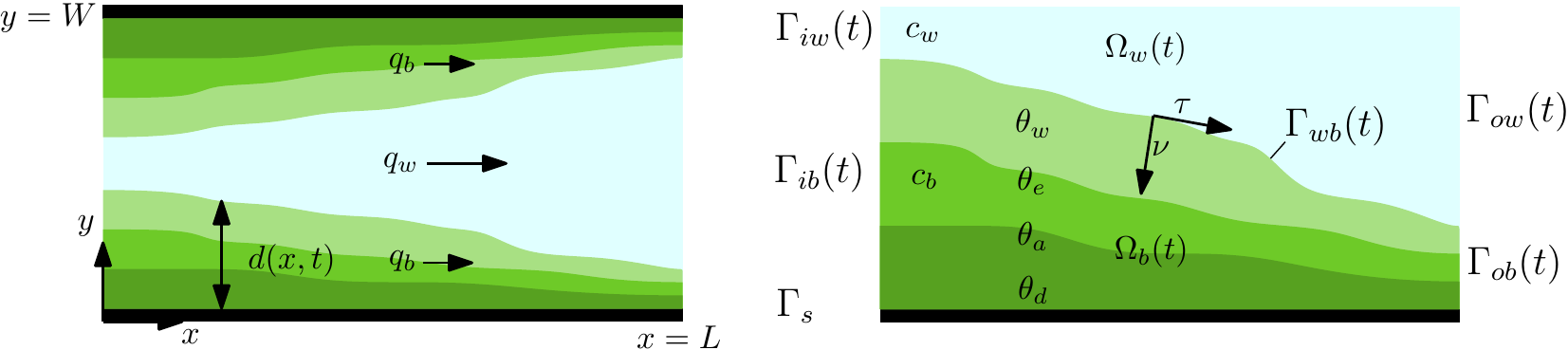}
%
\caption{The strip, coordinate system and biofilm thicknesses (left). Zoom into the domain close to the strip boundary (right).}
\label{domains}       
\end{figure}

We consider a simplified domain, a two-dimensional pore of length $L$ and width $W$, denoted by $\Omega:=(0,L)\times (0,W)$. The biofilm has four components: water, EPS, active, and dead bacteria ($j=\lbrace w,e,a,d\rbrace$). Let $\theta_j(t,\textbf{x})$ and $\rho_j$ denote the volume fraction and the density of species $j$. The biomass phases and water are assumed to be incompressible and that the biofilm layer is attached to the pore walls. Further the volumetric fraction of water is taken as constant $\partial_t\theta_w=0$. As observed in Fig. \ref{domains}, the biofilm layers has a thickness $d=d(x,t)$ that changes in time and depends on the location at the pore wall. For simplicity we assume symmetry in the y-direction. With this we distinguish the following:
\begin{align*}
\Omega_w(t):=&\lbrace (x,y)\in\mathbb{R}^2|\;x\in(0,L),\; y\in(d(x,t),W/2) \rbrace-\text{the water domain,}\\
\Omega_b(t):=&\lbrace (x,y)\in\mathbb{R}^2|\;x\in(0,L),\; y\in(0,d(x,t))\rbrace-\text{the biofilm layer,}\\
\Gamma_{wb}(t):=&\lbrace (x,y)\in\mathbb{R}^2|\;x\in(0,L),\;y=d(x,t) \rbrace-\text{the biofilm-water interface.}
\end{align*}
$\Gamma_{ib}$, $\Gamma_{iw}$, $\Gamma_{ob}$, and $\Gamma_{ow}$ are the inflow and outflow boundaries in the water and biofilm. $\Gamma_{s}$ is the pore wall boundary.
\\
The water flow is described by the Stokes model
\begin{equation}
\mu\Delta \textbf{q}_w=\nabla p_w,\quad\nabla\cdot \textbf{q}_w=0\quad\text{in}\;\Omega_w(t),
\end{equation}
where $\mu$ is the viscosity, $p_w$ is the water pressure, and $\textbf{q}_w$ is the water velocity. For the water flux in the biofilm one has the Darcy law and mass conservation
\begin{equation}
\nabla\cdot\textbf{q}_{b}=0,\quad\textbf{q}_{b}=-\frac{k\theta_w}{\mu}\nabla p_{b}\quad\text{in}\;\Omega_b(t),
\end{equation}
where $\textbf{q}_{b}$ and $p_{b}$ are the velocity and pressure of the water in the biofilm respectively, and $k$ is the permeability of the biofilm. At the interface, one has the water mass balance, the equilibrium of normal forces across $\Gamma_{wb}(t)$, and the well-known Beavers-–Joseph-–Saffman boundary condition \cite{Beavers:Article:1967,Saffman:Article:1971,Mikelic:Article:2000}
\begin{eqnarray}
(\textbf{q}_{w}-\textbf{q}_{b})\cdot\pmb{\nu}&=&\nu_n(1-\theta_w)\;\;\text{on}\;\Gamma_{wb}(t),\\
-\pmb{\nu}\cdot(-p_w\mathbb{1}+\mu(\nabla\textbf{q}_w+\nabla\textbf{q}^T_w))\cdot\pmb{\nu}&=&p_{b}\quad\quad\quad\quad\;\text{on}\;\Gamma_{wb}(t),\\
-\pmb{\nu}\cdot(-p_w\mathbb{1}+\mu(\nabla\textbf{q}_w+\nabla\textbf{q}^T_w))\cdot\pmb{\tau}&=&\frac{\alpha}{\sqrt{k}}\textbf{q}_{w}\cdot\pmb{\tau}\quad\text{on}\;\Gamma_{wb}(t).
\end{eqnarray}
Here $\alpha$ is the Beavers–-Joseph constant, $\nu_n$ is the normal velocity of the interface, $\pmb{\nu}$ is the unit normal pointing into the biofilm, and $\pmb{\tau}$ is the unit tangential vector,
\begin{equation}
\pmb{\nu}=(\partial_x d,-1)^T/\sqrt{1+(\partial_x d)},\quad\pmb{\tau}=(1,\partial_x d)^T/\sqrt{1+(\partial_x d)}.
\end{equation}
For each biofilm component $(l\in\lbrace e,a,d\rbrace)$ one has mass conservation \cite{Alpkvist:Article:2007}
\begin{equation}
\frac{\partial\rho_l\theta_l}{\partial t}+\nabla\cdot(\textbf{u}\rho_l\theta_l)=R_l\quad\text{in}\;\Omega_b(t)
\end{equation}
where $R_l$ are the rates on the biomass volume fractions, $\rho_l$ is the density of the $l$ component, and $\textbf{u}$ is the velocity of the biomass. Following \cite{Duddu:Article:2009} one has
\begin{equation}
\textbf{u}=-\nabla\Phi\quad\text{in}\;\Omega_b(t),
\end{equation}
where $\Phi$ is the growth velocity potential satisfying
\begin{eqnarray}
-\nabla^2\Phi&=&\frac{1}{1-\theta_w}\sum_l\frac{R_l}{\rho_l}\quad\text{in}\; \Omega_b(t).
\end{eqnarray}
We adopt Monod-type reaction rates
\begin{eqnarray}
R_a&=&Y_a\mu_n\theta_a\rho_a\frac{c_{b}}{K+c_{b}}-k_{\text{res}}\theta_a\rho_a,\\
R_e&=&Y_e\mu_n\theta_a\rho_a\frac{c_{b}}{K+c_{b}},\\
R_d&=&k_{\text{res}}\theta_a\rho_a,
\end{eqnarray}
where $Y_e$ and $Y_a$ are yield coefficients, $k_{\text{res}}$ is the decay rate, and $K$ is the Monod half nutrient velocity coefficient.

Denoting by $c_w$ and $c_b$ the nutrient concentration in the water and biofilm respectively, one has
\begin{eqnarray}
\partial_t c_w+\nabla\cdot\pmb{J}_w&=&0,\;\; \pmb{J}_w=-D\nabla c_w+\textbf{q}_wc_w\quad\;\;\;\;\text{in}\;\Omega_w(t),\\
\partial_t(\theta_w c_b)+\nabla\cdot\pmb{J}_b&=&R_b,\;\;\pmb{J}_b=-\theta_wD\nabla c_b+\textbf{q}_{b}c_{b}\quad\;\text{in}\;\Omega_b(t),
\end{eqnarray}
where $D$, $\pmb{J}_w$, $\pmb{J}_b$, and $R_b$ are the nutrient diffusion coefficient in water, the nutrient flux in the water and biofilm, respectively, and the nutrient reaction term. For the latter one has
\begin{equation}
R_b=-\mu_n\theta_a\rho_a\frac{c_{b}}{K+c_{b}}
\end{equation}
where $\mu_n$ is the maximum rate of nutrient. The coupling conditions are given by the Rankine-Hugoniot condition and the continuity of the concentrations:
\begin{eqnarray}
(\pmb{J}_b-\pmb{J}_w)\cdot \pmb{\nu}&=&\nu_n(\theta_wc_b-c_w)\quad\text{on}\;\Gamma_{wb}(t),\\
\theta_wc_{b}&=&c_{w}\quad\quad\quad\quad\quad\quad\text{on}\;\Gamma_{wb}(t).
\end{eqnarray}
Accounting for the biofilm dettachment due to shear stress, the normal velocity of the interface is given by \cite{vanNoorden:Article:2010}
\begin{equation}
\nu_n=\begin{cases}
[\pmb{\nu}\cdot\vec{u}]_{+},& d=W,\\
\pmb{\nu}\cdot\vec{u}+k_{str}\mu||(\mathbb{1}-\pmb{\nu}{\pmb{\nu}}^T)(\nabla \textbf{q}_w+\nabla \textbf{q}_w^T)\pmb{\nu}||,&0<d<W,\\
0,&d=0,\end{cases}
\end{equation}
where $k_{srt}$ is a constant for the shear stress and
\begin{equation}
\nu_n=-\frac{\partial_t d}{\sqrt{1+(\partial_xd)^2}}.
\end{equation}
\subsection{Boundary and initial conditions}
At the inflow we specify the pressure and nutrient concentration and we consider homogeneous Neumann condition for the growth velocity potential and volumetric fractions. At the outflow, we specify the pressure and take Neumann conditions for the concentrations, growth velocity potential, and volumetric fractions. At the substrate, we choose no-flux for the water, nutrients, and volumetric fractions and homogeneous Neumann conditions for the growth potential. We prescribe slip boundary condition for the water flux and homogeneous Neumann conditions for the nutrients. At the biofilm-water interface, we set the growth potential to zero and we consider homogeneous Neumann conditions for the volumetric fractions. The model is completed by the given initial data for the pressure, volumetric fractions, biofilm thickness, growth potential, and nutrient concentration.
 
\section{Implementation}
The challenging aspects in the numerical solution of the mathematical model are due to the coupling of the non-linearities and the existence of a free boundary (the biofilm-water interface) as it needs to be determined as part of the solution process. We use an ALE method for tracking the free boundary \cite{Donea:Article:2004}. This method offers an accurate representation of the interface and allows to write the mass transport equations across the interface without any modification. However, the mesh will become time dependent.

\begin{figure}
\includegraphics[scale=1]{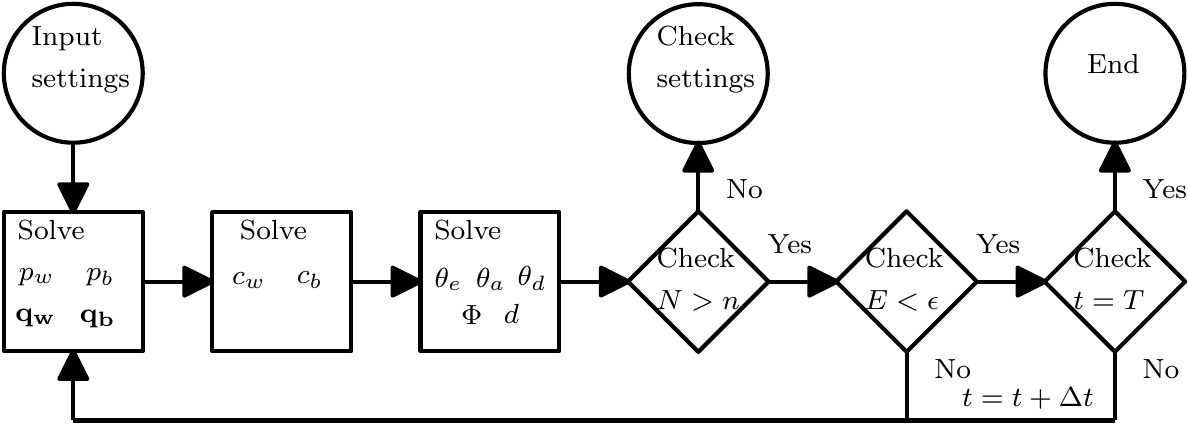}
%
\caption{Flow tree for solving the mathematical model.}
\label{flow}       
\end{figure}

There are plenty of approaches for solving coupled system of partial differential equations; for example, implicit iterative methods \cite{List:Article:2016,Pop:Article:2004,Radu:Article:2015}. We use backward Euler for the time discretization and linear Galerkin finite elements for the discretization in space. To solve the system of equations, we split the solution process into three steps: the flux ($p_w,\pmb{q}_w,p_b,\pmb{q}_b$), the nutrient ($n_w,n_b$), and the biofilm ($\theta_e,\theta_a,\theta_d,\phi,d$) variables. Firstly, a damped version of Newton's method is used to find the solution of the flux variables considering the initial conditions of all variables. Secondly, we find the solution of the nutrient variables considering the previous solution of the flux variables and initial conditions of the remaining variables. After, we find the solution of the biofilm variables considering the previous solution of the flux and nutrient variables and initial conditions of the biofilm variables. We iterate between the three steps until the error $E$ (the difference between successive values of the solution) drops below a given tolerance $\epsilon$. If the error criterion ($E<\epsilon$) is satisfied before $N$ iterations ($n<N$), we move to the next time step and solve again until a given final time $T$. Fig \ref{flow} shows the flowchart for the computational scheme. The infinity norm is used for the vector norm computation in the shear stress equation. The model equations are implemented in the commercial software COMSOL Multiphysics (COMSOL 5.2a, Comsol Inc, Burlington, MA, www.comsol.com).

\section{Numerical simulation of biofilm formation in a strip}
\begin{figure}[h!]
\includegraphics[scale=.4]{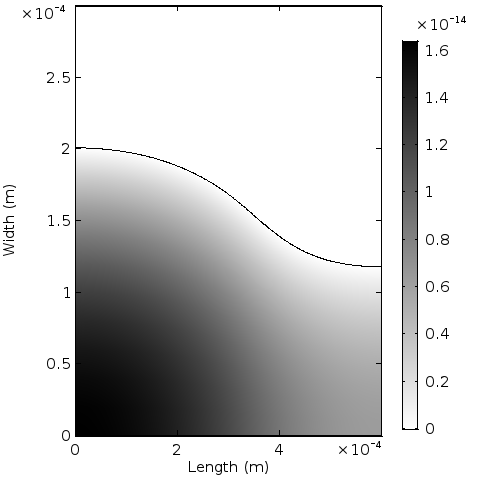}\quad\includegraphics[scale=.4]{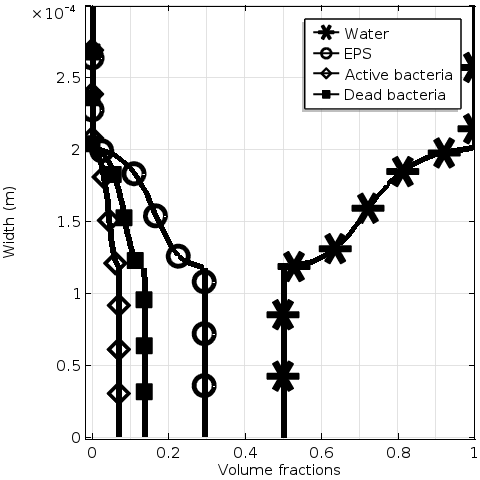}
\caption{Growth velocity potential (left) and total volume fractions (right).}
\label{author_mini4:fig:2}       
\end{figure}

In Table \ref{Landa:tab:1}, the values of parameters for the numerical simulations are presented. These values are taken from \cite{Alpkvist:Article:2007}, where the authors built a model for heterogeneous biofilm development. The dimensions of the strip and the injected nutrient concentration are also taken from  \cite{Alpkvist:Article:2007}, with values $L=600\times 10^{-6}m$ and $W=600\times 10^{-6}m$. From previous studies, we set the biofilm permeability $k=1\times 10^{-10}m^2$ \cite{Deng:Article:2013}, the Beavers--Joseph constant $\alpha=0.1$ \cite{Beavers:Article:1967}, and the stress coefficient $k_{str}=9\times 10^{-11}\;m/(s\;Pa)$ \cite{Landa-Marban1:Article:2017}. 
%
\begin{table}
\caption{Table of model parameters for the verification study \cite{Alpkvist:Article:2007}}
\label{Landa:tab:1}       
%
%
\begin{tabular}{p{1.3cm}p{2.2cm}p{1.3cm}p{2cm}p{1.3cm}p{2.6cm}}
\hline\noalign{\smallskip}
Symbol & Value & Symbol & Value& Symbol & Value\\
$\mu_n$ &	$1.1\times 10^{-5}/\text{s}$&$\rho_w$	& 	$1000\;\text{kg}/\text{m}^{3}$&$K$	& 	$1\times 10^{-4}\;\text{kg}/\text{m}^{3}$\\
$k_\text{res}$ & 	$2\times 10^{-6}/\text{s}$&$\rho_e$	& 	$60\;\text{kg}/\text{m}^{3}$&$D$	& 	$9.6\times 10^{-10}\;\text{m}^2/\text{s}$\\
$Y_{a}$ & 	$.22$&$\rho_a$	& $60\;\text{kg}/\text{m}^{3}$&$\mu$	& 	$1\times 10^{-3}\;\text{Pa}\cdot\text{s}$\\
$Y_{e}$	& $.4$&$\rho_d$	& 	$60\;\text{kg}/\text{m}^{3}$\\ 
\noalign{\smallskip}\hline\noalign{\smallskip}
\end{tabular}
\end{table}\\
The initial biofilm thickness is $d_0=30\times 10^{-6}m$. The volume fraction of water in the biofilm is 50\%. The left half of the biofilm ($0<x<L/2$) consists of 50\% active bacteria, while the other half consists of 25\% active bacteria and 25\% EPS. Nutrients are injected at a pressure of $P=0.5\;\text{Pa}$ and a concentration of $C=0.001\;\text{kg}/\text{m}^{3}$. We run the numerical simulations for 300 hours. In Fig. \ref{author_mini4:fig:2}, the growth velocity potential and volumetric fractions are shown. The biofilm has a greater active bacterial volume fraction on the side where the nutrients are injected, leading to a greater biofilm growth in comparison with the biofilm on the right hand side. After 300 hours, we observe that the lower part of the biofilm is approximately 92\% formed by water, EPS, and dead bacteria while only 8\% is formed by active bacteria.

\subsection{Parametric studies}

\begin{figure}
\includegraphics[scale=.4]{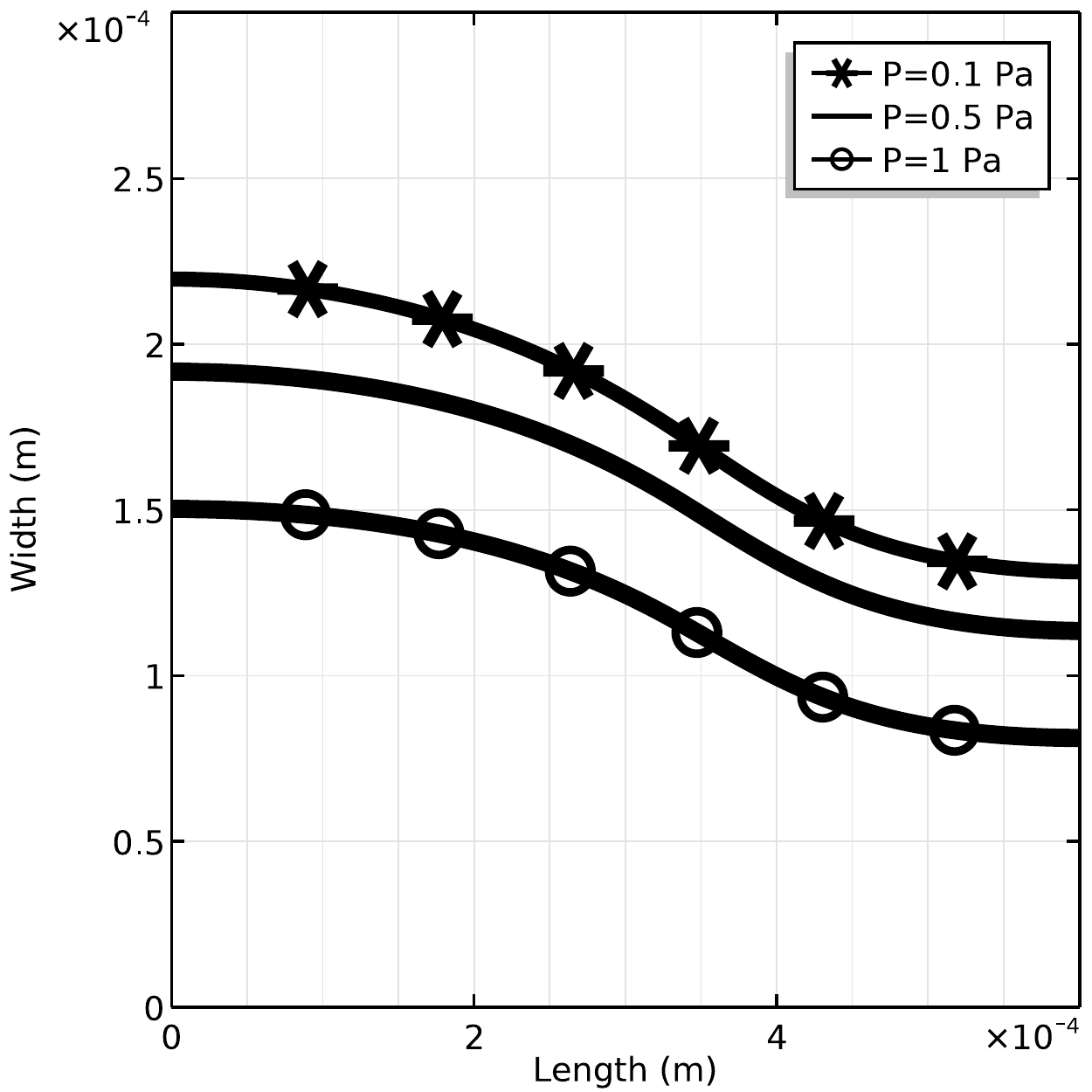}\quad\includegraphics[scale=.4]{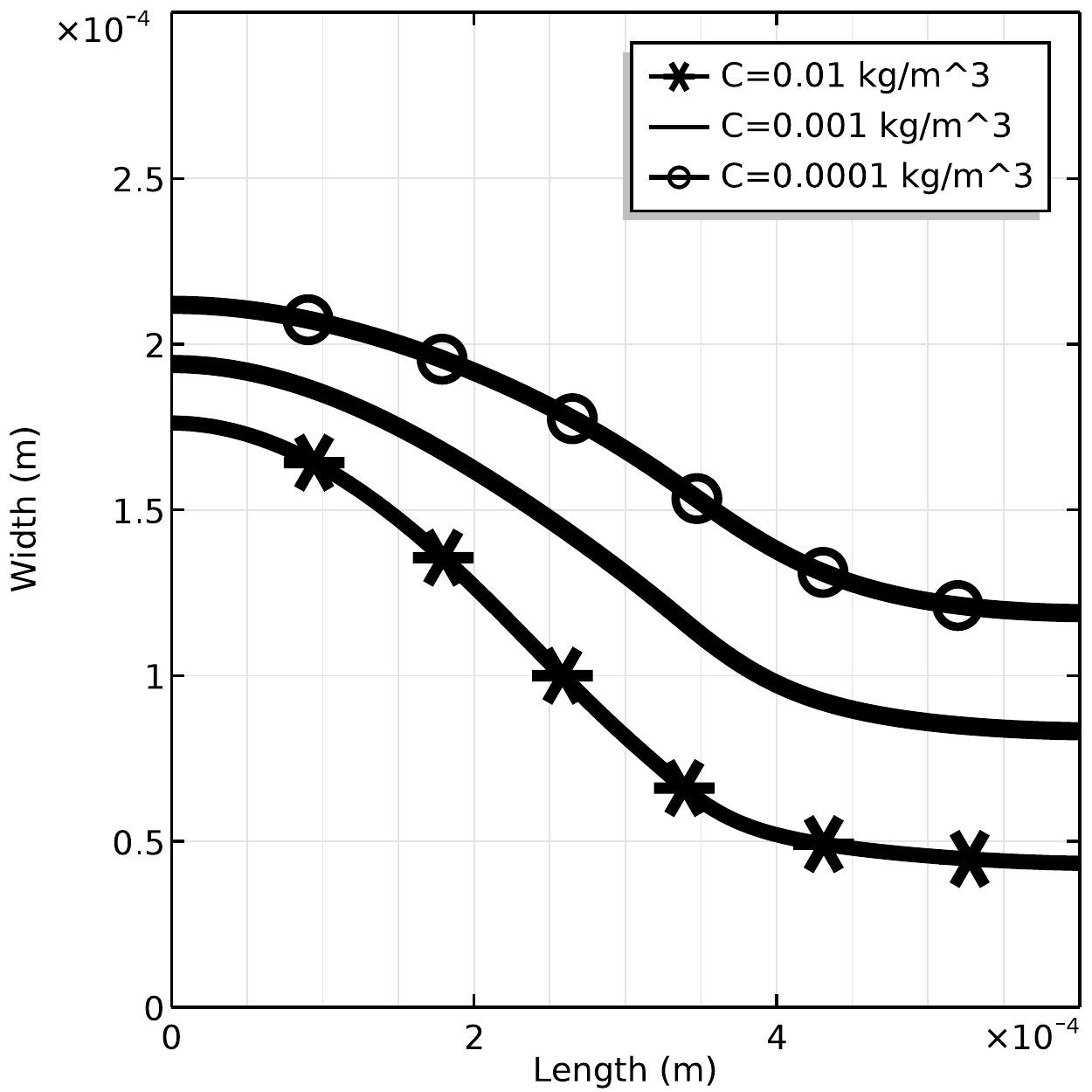}
\caption{Biofilm height at various pressures (left) and nutrient concentrations (right).}
\label{pressurenutrient}       
\end{figure}

Fig. \ref{pressurenutrient} shows the sensitivity analysis for the inlet pressure and injected nutrients. We observe that as the pressure increases, the biofilm growth is less due to the shear force. Regarding the nutrient concentration, we notice that increasing the nutrient concentration by a factor of 10, the biofilm has a faster growth. On the other hand, if we decrease the nutrient concentration by a factor of 10, the biofilm grows slowly and there is almost no biofilm growth on the outflow part.

\begin{figure}[h]
\includegraphics[scale=.4]{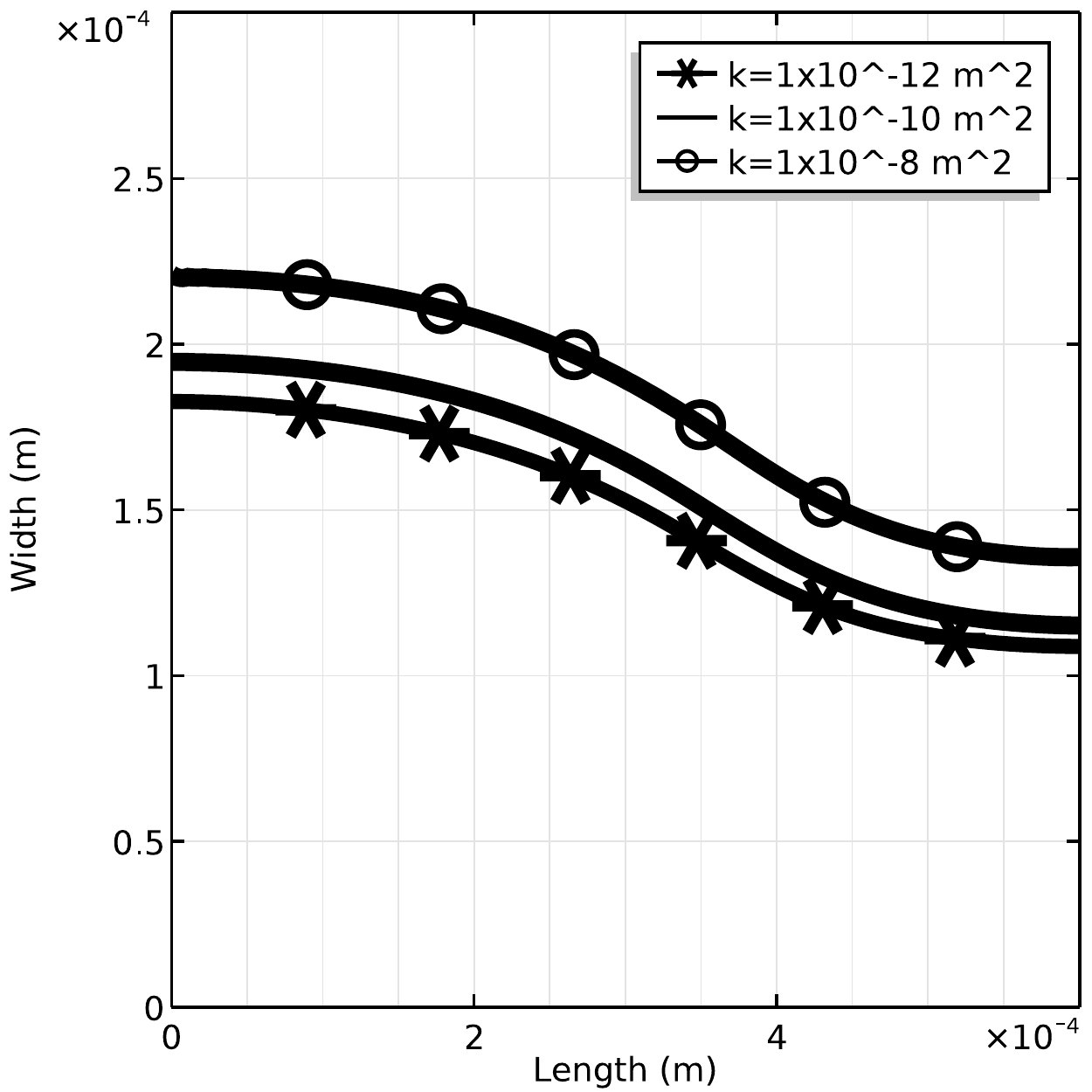}
\caption{Comparison of the permeable effects on the dynamic biofilm development.}
\label{compadre}       
\end{figure}

In order to quantify the impact of the biofilm permeability, we perform numerical simulations with different values of $k$. For the high permeable biofilm, we set $k=1\times 10^{-8}m^2$ while for the less permeable biofilm we set $k=1\times 10^{-12}m^2$.  We run again the numerical simulations for 300 hours. Fig. \ref{compadre} shows the modeled spatial distribution of biofilm height for the different permeabilities. We observe that for the less permeable biofilm, the biofilm formation is slower. The difference in the results is related to two different phenomena. First, the nutrient transport inside the less permeable biofilm is dominated by diffusion, while the transport in the high permeable biofilm is also due to convection. The second reason is that the coupling condition at the interface results in greater water velocity gradient for the impermeable biofilm, therefore the shear stress is larger.
\section{Conclusions}
In this work we have considered a mathematical model for permeable biofilm in a microchannel. The solution algorithm used for numerical simulations for the considered model has been presented. A sensitivity analysis is performed. We remark that if the flow rate is low, then the flux inside the biofilm can be neglected. However, for higher flow rates we must consider the effects of the flow inside the biofilm, that affects the transport of nutrients and the flux velocity value at the interface. The latter influence the biofilm thickness via the stress force.\\

\noindent \textbf{Acknowledgments}: The work of DLM and FAR was partially supported by the Research Council of Norway through the projects IMMENS no. 255426 and CHI no. 255510. ISP was supported by the Research Foundation-Flanders (FWO) through the Odysseus programme (G0G1316N) and Statoil through the Akademia grant.

\end{document}